\documentclass{IEEEtran}

\usepackage{amsmath, amssymb, amsthm}
\usepackage{mathptmx}
\DeclareMathSizes{11}{19}{13}{9}
\usepackage{tabularx}
\usepackage{booktabs}
\usepackage{algorithm, algorithmic}
\usepackage{url}
\usepackage[square,sort,comma,numbers]{natbib}
\usepackage{caption}
\usepackage[margin=1cm]{subcaption}
\usepackage{bbm}
\usepackage{wrapfig}
\usepackage{graphicx}
\newcommand{\prob}{\mathbb{P}}
\newcommand{\mgf}[1]{\mathbf{M}_{#1}}
\newcommand{\expec}{\mathbf{E}}

\newcommand{\naturals}{\mathbb{N}}

\newcommand{\graphrandom}{\mathcal{G}}
\newcommand{\partition}{\mathcal{C}}

\newcommand{\alter}{\mathfrak{A}}
\newcommand{\alterfun}{f}
\newcommand{\altRandGraph}{\Tilde{\mathcal{G}}}
\newcommand{\Graph}{G}

\newcommand{\RandGraph}{\mathcal{G}}
\newcommand{\VertSet}{V}

\newcommand{\EdgeSet}{E}

\newcommand{\subto}{\textrm{s.t.}}

\newcommand{\opt}{\textrm{OPT}}

\newcommand{\Fcal}{\mathcal{F}}

\newcommand{\defeq}{:=}

\newtheorem{theorem}{Theorem}

\newtheorem{lemma}[theorem]{Lemma}

\begin{document}

\title{Countering Misinformation on Social Networks Using Graph Alterations}

\author{Yigit Ege Bayiz, Ufuk Topcu \thanks{Y. E. Bayiz, and U. Topcu are with Univeristy of Texas at Austin, TX 78712 USA. E-mails: \{egebayiz, utopcu\}@utexas.edu.}}

\maketitle

\begin{abstract}
  We restrict the propagation of misinformation in a social-media-like environment while preserving the spread of correct information. We model the environment as a random network of users in which each news item propagates in the network in consecutive cascades. Existing studies suggest that the cascade behaviors of misinformation and correct information are affected differently by user polarization and reflexivity. We show that this difference can be used to alter network dynamics in a way that selectively hinders the spread of misinformation content. To implement these alterations, we introduce an optimization-based probabilistic dropout method that randomly removes connections between users to achieve minimal propagation of misinformation. We use disciplined convex programming to optimize these removal probabilities over a reduced space of possible network alterations. We test the algorithm's effectiveness using simulated social networks. In our tests, we use both synthetic network structures based on stochastic block models, and natural network structures that are generated using random sampling of a dataset collected from Twitter. The results show that on average the algorithm decreases the cascade size of misinformation content by up to $70\%$ in synthetic network tests and up to $45\%$ in natural network tests while maintaining a branching ratio of at least $1.5$ for correct information.

\end{abstract}


\section{Introduction}
Be it a deliberate spread of controversy caused by a disinformation campaign, or benign misinformation content that cascades through the internet, the propagation of false news is a major issue in social media networks. The increasing public consumption of social media over the last decade has caused more and more people to rely on social media as a source of news\citep{Edson2018SMnews, Marchi2012SMNews}. And the attempts to counter misinformation using manual content classification and human moderators have failed to scale up the sheer amount of content \citep{Gilbert2013reddit} that propagates through modern social networks. Therefore over the last decade, automated means of countering false news have drawn great interest.

Existing automated counters to false news mainly focus on the detection of misinformation content. The exact form of these detection algorithms depends on the type of content and the underlying social media network. In general, most misinformation detection methods rely on content classification using some black-box machine learning algorithm that is trained on a dataset labeled by humans. These content classification methods can yield high accuracy. However, these content classifiers still suffer from large biases caused by the biases in the training datasets \cite{Hirelkar2020nlpChallanges}. And their use in social media platforms can lead to ethnic, religious or political discrimination within the platform.

In this work, we deviate from the existing literature that mainly focuses on misinformation detection, and instead, we approach the problem of countering misinformation as an optimal control problem on a network. In this approach, we study the problem of altering the social network dynamics in a way that restricts misinformation spread while keeping the propagation of true content above some acceptable level. As such, we represent the problem of countering misinformation as an optimization problem on a social network model and then solve this optimization problem to find real-world changes to the social network that reduce the misinformation spread over the network.

In our social network model, we represent the propagation of news on a social media platform using a percolation process model. In this model, each news item propagates in a small-world network of users in consecutive cascades. In each cascade, the users in the network that believe a news content is correct re-share the news content probabilistically. Then, each user that receives the shared news content, either believes in it or discards it based on a probability distribution determined by the polarization of the sharing and receiving users and the reflexivity of the receiving user. 

The evidence suggests that there are subtle yet detectable differences in the cascade behaviors of misinformation and true content. User polarization and reflexivity are thought to be the main drivers of this difference \cite{vicario2016spreading, pennycook2021psychology}. Thus, the above model can capture different propagation patterns of misinformation and true content. Our approach to countering misinformation relies on this difference in propagation patterns to discriminate between different content types. We exploit this propagation difference to design alterations on the network of users that selectively hinders the spread of misinformation containing news while maintaining acceptable propagation of true content.
    
The specific methods that can be used to control the content flow over the network vary significantly depending on the capabilities and the structure of the underlying online platform. We assume a social media environment that can acquire usage data from the users and can estimate the polarization of the user as well as the probability to reshare a particular news content. Under this assumption, we propose a method called the \textit{Dropout Method}. This method relies on selectively limiting the content flow over the network by the random omission of news items in a user's news feed with a predetermined \textit{dropout probability}. These probabilities must be set up to reduce misinformation spread while minimally affecting the spread of correct content. To achieve this discrimination between misinformation and true content, we let the dropout probabilities depend on the polarization of both the sharing and receiving users. As mentioned before, these quantities are known to affect misinformation and true content flow differently, thus they can be used to identify news shares that are likely to contain misinformation.


\subsection{Main Contributions}
We have two main contributions:
\begin{enumerate}
    \item In section 3, we develop an optimization-based approach to model the problem of countering misinformation through alterations in the social network structure.
    \item  In section 4, we seek approximate solitouns to the problem we develop in section 3, and ultimately develop an algorithm that can counter misinformation through alterations in the social network structure.
\end{enumerate}

\subsection{Related Work}
\subsubsection{Misinformation Propagation}
The existing works on misinformation propagation can be separated into two different categories. the first of these categories focuses on finding mathematical models that describe the propagation of misinformation content. Most of these models describe the propagation of misinformation using established epidemiological models. The epidemiological models that have been most widely used in modeling viral content are, susceptible-infected-susceptible (SIS) \cite{kimura2009efficient, jin2013epidemiological}, susceptible-infected-removed (SIR) \cite{zhao2013sir, wang2017sir}, and susceptible-exposed-infected-removed (SEIR) \cite{xia2015rumor,liu2017analysis}. All of these models describe misinformation propagation over a social network by classification the users on the social network to different groups and modeling how these groups evolve over time. In their work Raponi et. al. provides a comprehensive analysis of these epidemiological models and their use in modeling misinformation spread \cite{raponi2022fake}. In our work we use an SIR-based model for content propagation as it is a widely accepted model for modelling fake news and it is simple to analyze.

The second category of works on misinformation propagation focuses on discriminating misinformation spread from the spread of other content. In their work Zhao et. al. statistically show that the propagation patterns of fake news differ predictably form other content \cite{Zhao2020}.  Wu et. al. uses support vector machine classifiers to detect identify misinformation campaigns based on propagation patterns \cite{Wu2015}. There are also mixed approaches to misinformation detection that uses both automated content classifiers and propagation patterns to detect misinformation. Varol et. al. uses a supervised learning approach based on k-nearest neighbors classifiers that uses sentiment values and propagation patterns to identify promoted campaigns on social media \cite{varol2017early}. Our approach is similar to these works in the sense that we seek discrimination between misinformation and other content. However, our method does not attempt to explicitly identify misinformation.

\subsubsection{Countering Misinformation}
There are some existing works that attempt to limit the propagation of misinformation. In their work Fan et. al. proposes two models for multiple competing diffusion processes on network and investigates the problem of containing rumor spread on a competitive diffusion model \cite{fan2014maximizing}. Similarly Litou et. al. model the competition between misinformation and credible information on a network using a novel dynamic linear threshold model and investigate the problem of finding optimal set of users on a network to initiate the propagation of credible content \citep{litou, litou2017efficient}. More recently there has been work on refining this approach by considering location \cite{zhu2019location} or community \cite{lv2019community} structures of the underlying social network. Unlike these works we do not focus on minimizing the influence of misinformation by maximizing the influence of a competing diffusion model. Instead we focus on altering the social media dynamics in a way that passively reduces misinformation spread without requiring any competing credible content.

\section{Preliminaries}
\subsection{Random Graphs}
A \emph{random directed graph} \cite{bollobas1998random} is a tuple $\RandGraph = (\VertSet, [p_{ij}])$ composed of a set $V$ of vertices, a set $E \subseteq V\times V$ of edges, and a matrix of edge probabilities $p_{ij}$ that assigns probability to each edge. An instance of a random directed graph $\RandGraph$ is a directed graph $\Graph = (V,\EdgeSet')$, where $\EdgeSet' \subseteq \EdgeSet$ and $\prob((i,j)\in \EdgeSet') = p_{ij}$.

\subsection{Discrete-Time SIR Model on Graphs}
A discrete-time \emph{susceptible-infected-removed} (SIR) model \cite{hill2010infectious} is a contagion propagation model that is often used to model propagation of epidemics. In this model, the contagion spreads in a network over consecutive iterations. Given a random directed graph $\graphrandom = (\VertSet, [p_{ij}])$, at each time $t$, The SIR model first splits the vertex set $V$ into three time-dependent partitions composed of a susceptible set $S_t$, an infected set $I_t$, and a removed set $R_t$. At each time step, each infected node $i \in I_t$ spreads its infection to all susceptible nodes $j \in S_t$, with probability $p_{ij}$. That is,
\begin{equation}\label{dynamics_inf}
    \prob(j \in I_{t+1} | j \in S_t) = 1 - {\sum_{\substack{i \in I_t \\ i \neq j} }{(1 - p_{ij})}}.
\end{equation}
Each node that gets infected remains infected for exactly $m$ turns where $m$ is a known integer constant. After that, they get removed. That is, for all $i\in I_t$,
\begin{equation}\label{dynamics_rmv}
    i \in \begin{cases}
        R_{t + 1}, \quad i \in I_\tau,\; \forall \tau \in \{t-m \dots t\} \\
        I_{t + 1}, \quad \textrm{otherwise}
    \end{cases}
\end{equation}
In our analysis and for the remainder of this work we take $m = 1$. This restricted form of SIR model is equivalent to another commonly used content propagation model called the \emph{independent cascade process} \cite{kempe2003}.

In addition to being used extensively in epidemics research, SIR models also see significant use in modelling viral content spread over social networks.

\subsection{Stochastic Block Models}
A stochastic block model (SBM) is a random graph model with inbuilt communities. We use the notation $\graphrandom_{SBM}(\partition, [b_{uv}]_{uv})$ to refer to an SBM model generated by a finite partition $\partition = \{C_1, C_2, \dots C_k\}$ of the set of nodes $V$ and a $k\times k$ SBM matrix $[b_{uv}]_{uv}$. We define this SBM as the random graph $\graphrandom = (V, E, [p_{ij}]_{ij})$, where $V = \bigcup_{u=1}^k{C_u}$ and the edge probabilities $p_{ij}$ are given as
\begin{equation}
    p_{ij} = b_{uv}\quad \textrm{for all}\quad i \in C_u,\, j \in C_v.
\end{equation}

\section{Models and Problem Setup}
\subsection{The Social Media Setup}
Consider a Twitter-like social media environment that has  $N$ users. We call any user posts or news articles that occur in this social media as \emph{content}. We use the term \emph{true content} to describe a content that contains correct or scientific information, and we use the term \emph{false content} to describe any content which contains misinformation, disinformation or conspiracy.

Contents are spread over the social media environment through \emph{shares} between users. Once a user receives a content, they can freely choose whether to \emph{re-share} it again. A content always originates from a subset of users, which we call \emph{seeds}, and notate as $I_0$. In practice the number $|I_0|$ of seeds is almost always small compared to the total number $N$ of users.

Following the Twitter model, we assume that the information spreads over the network in \emph{multicast} fashion. That is, whenever a user shares a content, the content is transmitted to all \emph{receivers} simultaneously. We also assume that each user can share or re-share a particular content only once. 

In practice, user shares can occur at any time $t \in [0,\infty)$. However, modelling and analysis in this continuous time domain is difficult. Thus, in our network model we use the discretized time $t \in \naturals = \{0,1,2,3,...\}$. The propagation of content in the network can be summarized as the following iterative process.
\begin{enumerate}
    \item Set $t = 0$.
    \item Content originates form seeds $I_0$. The seeds share the content.
    \item Set $t \leftarrow t + 1$.
    \item Some of the users that receive the content decide to re-share it again.
    \item If there are new re-shares, return to step $3$.
\end{enumerate}

\subsection{Modelling Content Propagation}
Given a social network with $N$ users, let $\partition = \{C_1, C_2, \dots, C_k\}$ denote the partition on the set of users that is induced by user polarization. These partitions can be generated by the political or moral opinions of the users, as well as the echo chambers that exist over the network. We model each user's probability of re-sharing received content using two quantities:
\begin{itemize}
    \item $r_i^+$: The probability that the user $i$ re-shares a received true content.
    \item $r_i^-$: The probability that the user $i$  re-shares a received false content.
\end{itemize}
These quantities reflect both \emph{reflexivity}, which is the ability to discriminate true and false content, and the probability of each user re-sharing the content they believe to be true.

We model the probability of a content shared by user $i \in C_u$ to be received by user $j \in C_v$ as a constant probability $c^-_{uv}$ and $c^+_{uv}$ for false and true content respectively. In practice, for social media networks, we have $c_{uu}^- \geq c_{uv}^-$ and $c_{uu}^+ \geq c_{uv}^+$ for all $u \neq v$ since the echo chambers that result from the user polarizations encourage content sharing between agents within the same polarization class and discourage content propagation across different polarization classes. We can write the total probability of content being transferred from user $i \in C_u$ to $j \in C_v$ as
\begin{subequations}
\begin{align}
    p^-_{ij} &= r^-_i c^-_{uv} \quad \textrm{for false content},\\
    p^+_{ij} &= r^+_i c^+_{uv} \quad \textrm{for true content}.
\end{align}
\end{subequations}

Following the social media setup in section 3.1, we model the content propagation using an SIR model with an infectious period of $1$. Here the partition $S_t$ represents the users that have not yet received a piece of content at time $t$, $I_t$ represents the users that have received the content in the current time step $t$, and $R_t = \bigcup_{\tau = 1}^{t-1} I_\tau$ is the set of nodes that have previously received the content. 

We can approximate these content propagation dynamics as an SIR model with infectious period of $1$ on one of two stochastic block models:
\begin{subequations}
\begin{align}
    \graphrandom^- &\defeq \graphrandom_{SBM}(\partition, [b^-_{uv}]_{uv}) \quad \textrm{for false content},\\
    \graphrandom^+ &\defeq \graphrandom_{SBM}(\partition, [b^+_{uv}]_{uv}) \quad \textrm{for true content},
\end{align}
\end{subequations}
where for all $u,v \in \{1,2,\dots, k\}$ we have
\begin{subequations}
\begin{align}
    b^-_{uv} = \frac{1}{|C_u|}\sum_{i \in C_u} r^-_i c^-_{uv},\\
    b^+_{uv} = \frac{1}{|C_u|}\sum_{i \in C_u} r^+_i c^+_{uv}.
\end{align}
\end{subequations}

The difference between the stochastic block models $\RandGraph^+$ and $\RandGraph^-$ results in different content propagation characteristics to be predicted by the SIR model. This reflects the difference in content propagation patterns that can be seen between real-world true and false content. When correctly fitted to the actual social media network, these simplified content propagation models are often capable of sufficiently capturing the difference in propagation patterns between true and false content.

We assume throughout this work that we know $b^-_{uv}$ and $b^+_{uv}$. This is a reasonable assumption since we can fit these SBMs to the real data collected from social media by directly estimating $b^-_{uv}$ and $b^+_{uv}$. A simple method of doing this is by first observing the propagation patterns of sample contents which are known to be either true or false, and then using a frequentist estimation of $b^-_{uv}$ and $b^+_{uv}$ from the observed propagation patterns. As more propagation data on true/false contents become available, this estimation can be repeated periodically to refine the estimates for $b^-_{uv}$ and $b^+_{uv}$ over time.

This estimation process requires reliable knowledge of whether the observed sample contents are true or false. Therefore we require reliable content classification to fit the SBMs $\RandGraph^+$ and $\RandGraph^-$ to the actual social media. To minimize the bias and fairness issues associated with automated content classification systems, we suggest doing the content classification either using user responses to the content (likes, comments, etc.) or manually by expert human moderators. This is possible, since there is no strict requirement to classify content during its propagation period, and the classification can easily be done afterward without any constraint on time. As we elaborate in the following sections,  the fact that automated content classification is superfluous for estimating content propagation models is one of the major advantages of our approach.

\subsection{Graph Alterations and Dropouts}
To counter the spread of false content over the social media network we need to determine how we can control the content propagation over the network. The classical way of achieving this is first determining if a piece of content is true or false using automated algorithms, and then restricting, or banning the content which is determined to be false. This is an effective means of stopping the propagation of content that is classified as false. However, as mentioned previously, this approach suffers from its explicit reliance on automated content classification methods, and the issues this reliance brings.

To resolve this reliance on automated agents, we introduce a network-design-based approach to counter false content called \emph{graph alterations}. Let $\alterfun: V\times V \times [0,1] \rightarrow [0,1]$ be a function that given a node pair $i,j$ and content transfer probability $p_{ij}$, generates an altered content transfer probability of $\alterfun(i,j,p_{ij})$. We define a graph alteration $\alter_\alterfun: \RandGraph \mapsto \altRandGraph$ as the mapping induced by $\alterfun$ between two random graphs. That is for any random graph $\RandGraph \defeq (V,[p_{ij}]_{ij})$ we have 
\begin{equation}
    \alter_\alterfun(\RandGraph) \defeq \altRandGraph \defeq (V,[\alterfun(i,j,p_{ij})]_{ij}).
\end{equation}
In other words, given a random graph $\RandGraph$, with transfer probabilities $[p_{ij}]_{ij}$ $\alter_\alterfun(\RandGraph)$ is a new random graph with altered transfer probabilities $[\alterfun(i,j,p_{ij})]_{ij}$.

Suppose that we have two random graphs $\RandGraph^+$ and $\RandGraph^-$ describing the propagation of true and false content respectively on a social network. For any fixed $\alterfun$, applying the same graph alteration $\alter_\alterfun$ to $\RandGraph^-$ and $\RandGraph^+$ provides a method to alter the structure of both of these graphs simultaneously in a way that does not explicitly depend on the content type. Each graph alteration $\alter_\alterfun$ corresponds to a change in the social network structure that results in the true and false contents to propagate according to random graphs $\alter_\alterfun(\RandGraph^+)$ and $\alter_\alterfun(\RandGraph^-)$ respectively.

The set of graph alterations that are feasible to implement on a social media network depend heavily on the design and capabilities of the social media platform. In this work, we focus on the graph alterations corresponding to randomized content dropouts. We define a dropout as the artificial prevention of a content transfer between two agents. For example, in a Twitter-like social media platform, we can implement such randomized content dropouts by artificially excluding a content from the receiver's feed. In the SBM model described in Section 3.4, we model a random dropout between two users $i\in C_u$ and $j\in C_v$ using an altered content transfer probability function.
\begin{equation} \label{dropout_fun}
    f_d(i,j,p_{ij}) \defeq d_{uv} p_{ij},
\end{equation}
where $d_{uv}$ are the dropout probabilities. Then under graph alteration $\alter_{\alterfun_d}$, the altered SBMs corresponding to true and false content propagation graphs are 
\begin{subequations}
\begin{align}
    \altRandGraph^- \defeq \alter_{\alterfun_d}(\RandGraph^-)&\defeq \graphrandom_{SBM}(\partition, [\Tilde{b}^-_{uv}]_{uv}) \quad \textrm{for false content},\\
    \altRandGraph^+ \defeq\alter_{\alterfun_d}(\RandGraph^+) &\defeq \graphrandom_{SBM}(\partition, [\Tilde{b}^+_{uv}]_{uv}) \quad \textrm{for true content},
\end{align}
\end{subequations}
where
\begin{subequations}
\begin{align}
    \Tilde{b}^-_{uv} & \defeq d_{uv} b^-_{uv},\\
    \Tilde{b}^+_{uv} & \defeq d_{uv} b^+_{uv}.
\end{align}
\end{subequations}
Assuming that the original SBMs $\RandGraph^-$ and $\RandGraph^+$ are accurate models of content propagation over the network, we can use these altered SBMs to predict the effect of graph alterations on the real-world social network. 

\subsection{Problem Statement}
In the most general case, the problem of countering misinformation is finding a graph alteration $\alter_{\alterfun}$ that minimizes the predicted propagation of false content, whilst keeping the predicted propagation of true content above some acceptable level. That is for a given safety parameter $\alpha$, at every time $t$ we wish to solve the following optimization problem. 
\begin{subequations}
\label{eqn::general_problem}
\begin{align}
    \min_{\alter_{\alterfun}}&\quad \expec_{\altRandGraph^-}[|I_{t+1}| |S_{t}, I_{t}, R_{t}], \\
    \subto &\quad \expec_{\altRandGraph^+}[|I_{t+1}| |S_{t}, I_{t}, R_{t}] \geq \alpha |I_{t}|.
\end{align}
\end{subequations}

The general case given in \eqref{eqn::general_problem} is a non-convex optimization problem over all possible graph alterations $\alter_{\alterfun}$. This is intractable for large networks. However, we can reduce it into a simpler problem by restricting and parametrizing the graph alterations $\alter_{\alterfun}$. Throughout the rest of this work, we restrict our analysis to graph alterations $\alter_{\alterfun_d}$ that are generated by randomized dropouts $[d_{uv}]_{uv}$ where $d_{uv}$ is the dropout probability of a content transfer from a user in polarization class $C_u$ to a user in polarization class $C_v$. This yields the following optimization problem.
\begin{subequations}\label{eqn::dropout_problem}
\begin{align}
    \min_{d \in [0,1]^{k\times k}}& \quad \expec_{\alter_{\alterfun_d}(\RandGraph^-)}[|I_{t+1}| |S_{t}, I_{t}, R_{t}], \\
    \subto &\quad \expec_{\alter_{\alterfun_d}(\RandGraph^+)}[|I_{t+1}| |S_{t}, I_{t}, R_{t}] \geq \alpha |I_{t}|,    \label{eqn::dropout_problem_constraint}
\end{align}
\end{subequations}
where $\alterfun_d$ is the altered transfer probability function defined in eq. \eqref{dropout_fun}, and $\alter_{\alterfun_d}$ is the graph alteration induced by $\alterfun_d$. 

After solving these optimization problems with model social networks $\RandGraph^-$ and  $\RandGraph^+$, we use the optimal graph alterations found by these problems to alter the content transfer probabilities of the real-world social network. Since the optimal solutions of problems \eqref{eqn::dropout_problem} and \eqref{eqn::general_problem} are time-dependent, we need to update the graph alteration at each time $t$ by re-evaluating the optimal solution to the optimization problem at hand based on the observed $S_t, I_t, R_t$.

\section{Theory and Algorithms}
As a general solution template, we consider the dynamic false content minimization loop given in Algorithm \ref{alg::general}. We run Algorithm \ref{alg::general} independently for each content that propagates over the network. In Algorithm \ref{alg::general} $\opt( \textrm{problem \eqref{eqn::dropout_problem}})$ refers to the optimal solution of problem \eqref{eqn::dropout_problem}, and Observe$(S_{t+1}, I_{t+1}, R_{t+1}| S_t, I_t, R_t, \altRandGraph)$ returns the observed $S_{t+1}, I_{t+1}, R_{t+1}$ sets generated by an SIR model on altered social network $\altRandGraph$ with known current state $S_t, I_t, R_t$. That is, the $S_{t+1}, I_{t+1}, R_{t+1}$ denotes the next set of susceptible, infected, removed users given that the real social network is altered using a dropouts $d^*$. This dropout is the optimal dropout based on our model networks $\RandGraph^+, \RandGraph^-$, which are stochastic block models as described in the previous section. Intuitively, given a piece of content that propagates as an SIR model on a real-world network $\RandGraph$ with unknown transfer probabilities, Algorithm \ref{alg::general} attempts to minimize the spread of the content if it propagates like a false content, and preserves the spread of the content if it propagates like a true content.
\begin{algorithm}
    \caption{False Content Minimization}
    \begin{algorithmic}[1]
    \REQUIRE Model Networks $\RandGraph^-, \RandGraph^+$, 
    Real-World Network $\RandGraph$, 
    Set of seed users $I_0$, 
    Safety parameter $\alpha$.
    \STATE $t \gets 0$
    \STATE $S_0 \gets V \setminus I_0$
    \STATE $R_0 \gets \emptyset$
    \WHILE{$|I_t| > 0$}
    \STATE $d^* = \opt( \textrm{problem \eqref{eqn::dropout_problem}})$
    \STATE $\altRandGraph \gets \alter_{\alterfun_{d^*}} (\RandGraph)$ \COMMENT {Alter the real-world social network.}
    \STATE $S_{t+1}, I_{t+1}, R_{t+1} \gets$ Observe$(S_{t+1}, I_{t+1}, R_{t+1}| S_t, I_t, R_t, \altRandGraph)$
    \STATE $t \gets t + 1$
    \ENDWHILE
    \end{algorithmic}
    \label{alg::general}
\end{algorithm}

The problem given in eq. \eqref{eqn::dropout_problem} is a non-convex problem. To solve it, we formulate an asymptotic approximation of it by considering the behavior of $\altRandGraph^+ = \expec_{\alter_{\alterfun_d}(\RandGraph^+)}[|I_{t+1}| |S_{t}, I_{t}, R_{t}]$ as the number $N$ of users diverges towards infinity. This is a reasonable approximation since the number $N$ of users in real social networks is often large enough that the inaccuracies caused by the asymptotic approximation is negligible compared to other sources of model inaccuracy.

We define $I^u_t = I_t \cap C_u$ as the set of infected users in polarization class $C_u$ at iteration $t$. Similarly, we also let $R^u_t = R_t \cap C_u$, and $S^u_t = S_t \cap C_u$. Then, for any $j \in C_v$ we have
\begin{equation} \label{eqn: false_prob_exact}
    \prob_{\altRandGraph^+}[j \in I_{t + 1}| S_{t}, I_{t}, R_{t}] = \begin{cases}
        1 - \prod_{u = 1} ^k (1 - d_{uv}b^+_{uv})^{|I^u_t|},  &\; j \in S_{t},\\
        0, &\;j \not \in S_{t}.
    \end{cases}
\end{equation}
This probability immediately follows from the transition probabilities of the altered SBM $\altRandGraph^+ = \alter_{\alterfun_d}(\RandGraph^+)$ corresponding to the false content.

The social networks that we are interested in often have a large number of users. Thus, we are interested in the asymptotic behavior of eq. \eqref{eqn: false_prob_exact} for large $N$. For such large networks we can approximate \eqref{eqn: false_prob_exact} as
 \begin{equation} \label{eqn: false_prob}
    \prob_{\altRandGraph^+}[j \in I_{t + 1}| S_{t}, I_{t}, R_{t}] =
    \begin{cases}
        1 - \prod_{u = 1} ^k \exp ( -|I^u_t| d_{uv}b^+_{uv} ),  &\, j \in S_{t},\\
        0, &\, j \not \in S_{t}.
    \end{cases}
\end{equation}
Under this approximation, for true content we have 
\begin{equation}\label{trueExpec}
    \expec_{\altRandGraph^+}[|I_{t + 1}|\,| S_{t}, I_{t}] =
        \sum_{v = 1}^k |S^v_{t}|\Big(1 - \exp\Big(-\sum_{u = 1} ^k|I^u_t| d_{uv}b^+_{uv} \Big)\Big),
\end{equation}
and similarly, for false content, we write 
\begin{equation}\label{falseExpec}
    \expec_{\altRandGraph^-}[|I_{t + 1}|\,| S_{t}, I_{t}] =
        \sum_{v = 1}^k |S^v_{t}|\Big(1 - \exp\Big(-\sum_{u = 1} ^k|I^u_t| d_{uv}b^-_{uv} \Big)\Big).
\end{equation}

Using equations \eqref{trueExpec} and \eqref{falseExpec}, we can rewrite the optimization problem given in \eqref{eqn::dropout_problem} as
\begin{subequations}
\label{eqn::dropout_problem_asymptotic}
\begin{align}
    \min_{d \in [0,1]^{k\times k}} &\quad \sum_{v = 1}^k |S^v_{t}|\Big(1 - \exp\Big(-\sum_{u = 1} ^k|I^u_t| d_{uv}b^-_{uv} \Big)\Big), \\
    \subto &\quad \sum_{v = 1}^k |S^v_{t}|\Big(1 - \exp\Big(-\sum_{u = 1} ^k|I^u_t| d_{uv}b^+_{uv} \Big)\Big) \geq \alpha |I_t|.
\end{align}
\end{subequations}

There is no guarantee that the optimization problem in eq. \eqref{eqn::dropout_problem_asymptotic} is feasible. In fact, since the left hand side of the constraint \eqref{eqn::dropout_problem_asymptotic} is monotonically increasing with $d_{uv}$ for all $u, v \in  \{1,\dots,k\}$ the problem \eqref{eqn::dropout_problem_asymptotic} is feasible if and only if we have 
\begin{equation}
    \sum_{v = 1}^k |S^v_{t}|\Big(1 - \exp\Big(-\sum_{u = 1} ^k|I^u_t| b^+_{uv} \Big)\Big) \geq \alpha |I_t|.
\end{equation}
That is, the problem \eqref{eqn::dropout_problem_asymptotic} is feasible exactly when  $(I_t)_t$  has branching ratio greater than or equal to $\alpha$ in the SIR model defined on the non-altered true content graph $\RandGraph^+$. Therefore choosing $\alpha$ too large can lead to infeasibility.

Another issue to note is that due to the dynamics of the SIR model given in equations \eqref{dynamics_inf} and \eqref{dynamics_rmv}, whenever we have $|I_\tau| = 0$ for some $\tau$, we guarantee $|I_t| = 0$ for all $t > \tau$. This means that even though \eqref{eqn::dropout_problem_asymptotic} might be feasible, the propagation of true content can halt if a random event leads to $|I_\tau| = 0$ at some $\tau$. Clearly, the probability of this event $|I_\tau| = 0$ decreases with larger safety parameter $\alpha$. But as stated previously, too large of a choice for the parameter $\alpha$ leads to infeasibility in the problem \eqref{eqn::dropout_problem_asymptotic}. Therefore, when choosing a safety parameter $\alpha$ one needs to consider a trade-off between feasibility, and robustness to probabilistic effects. In Lemma \ref{lemma1} we investigate this trade-off further and characterize the relation between $\alpha$ and the probability that the $|I_\tau| = 0$ given that the problem given in  \eqref{eqn::dropout_problem_asymptotic} is feasible. 

\begin{lemma} \label{lemma1}
    Suppose that there exists some $T \in \naturals$ such that the optimization problem \eqref{eqn::dropout_problem} is feasible for all $t \in [0,T]$, and let $(S_t, I_t, R_t)_t$ be the stochastic SIR process generated as in Algorithm \ref{alg::general}. Then if $\RandGraph = \RandGraph^+$, then for any non-negative $\lambda$ we have,
\begin{equation}
    \prob[\inf_{0\leq t\leq T}{|I_t|} = 0] \leq \expec\Big[e^{-\lambda \frac{|I_T|}{\alpha^T}}\Big] = \mgf{|I_T|}\Big(-\frac{\lambda}{\alpha^T}\Big),
\end{equation}
    where $\mgf{|I_T|}$ denotes the moment generating function of $|I_T|$.
\begin{proof}
    Let $(\Fcal_t)_{0\leq t \leq T}$ be the natural filtration generated by the stochastic process $(|I_t|)_{0\leq t \leq T}$. Let $Y_t = \frac{|I_t|}{\alpha^t}$. Then by the constraint of problem \eqref{eqn::dropout_problem} for all $t \in (0,T]$ we have $\expec [Y_{t+1} | \Fcal_t] = Y_t$. That is, $(Y_t)_t$ is Martingale. Then by Jensen's inequality, for all non-negative $\lambda$ and $t \in (0,T]$, we have $\expec [e^{-\lambda Y_{t+1}} | \Fcal_t] \geq e^{-\lambda Y_t}$. 
    Notice that $(\Fcal_t)_{t}$ is also a natural filtration for the stochastic process $(e^{-\lambda Y_t})_t$ since $e^{-\lambda Y_t}$ relates bijectively to $|I_t|$.
    Therefore $(e^{-\lambda Y_t})_t$ is a sub-Martingale sequence. Then,
    \begin{subequations}
    \begin{align}
        \prob[\inf_{0\leq t\leq T}{|I_t|} = 0] &= \prob[\inf_{0\leq t\leq T}{|I_t|} \leq 0]\\ 
        &= \prob[\inf_{0\leq t\leq T}{Y_t} \leq 0] \\
        &= \prob[\sup_{0\leq t\leq T}{e^{-\lambda Y_{t}}} \geq 1] \\
        &\leq \expec\Big[e^{-\lambda \frac{|I_T|}{\alpha^T}}\Big],
    \end{align}
    \end{subequations}
    where the last line follows from Doob's Martingale inequality.
\end{proof}
\end{lemma}

The optimization problem \eqref{eqn::dropout_problem_asymptotic} can be solved using gradient-based methods. However, it is also possible to simplify it further. We are mainly interested in the initial period of the viral spread of the content. That is, we want to counter false content before it spreads to a significant fraction of users. Similarly for true content, if we can ensure that the spread of true content is not restricted in the first couple of iterations, it is likely that a significant fraction of users will eventually receive the true content. Thus in practical applications, $N$ is usually much larger than $I_{t}$. Under this assumption, the inequalities
\begin{subequations}
\begin{align}
    \exp\Big(\sum_{u = 1} ^k|I^u_t| d_{uv}b^-_{uv} \Big) &\geq 1 - \sum_{u = 1} ^k|I^u_t| d_{uv}b^-_{uv}, \\
    \exp\Big(\sum_{u = 1} ^k|I^u_t| d_{uv}b^+_{uv} 
    \Big) &\geq 1 - \sum_{u = 1} ^k|I^u_t| d_{uv}b^+_{uv},
\end{align}
\end{subequations}
becomes tight. Therefore in the regime $N \>> I_{t}$ we can approximate the solution the optimization problem \eqref{eqn::dropout_problem_asymptotic} using the following optimization problem,
\begin{subequations}
\label{eqn::dropout_problem_asymptotic_linear}
\begin{align}
    \min_{d \in [0,1]^{k\times k}} &\quad \sum_{v = 1}^k \sum_{u = 1}^k |S^v_{t}| |I^u_t| d_{uv}b^-_{uv}, \\
    \subto &\quad \sum_{v = 1}^k \sum_{u = 1}^k |S^v_{t}|  |I^u_t| d_{uv}b^+_{uv} \geq \alpha |I_t|.
\end{align}
\end{subequations}
The above form is simply a linear program and it can be solved very efficiently using existing linear program solvers. As before, the optimization problem \eqref{eqn::dropout_problem_asymptotic_linear} is feasible if and only if we have
\begin{equation}
\label{feas_criterion}
    \sum_{v = 1}^k \sum_{u = 1}^k |S^v_{t}|  |I^u_t|b^+_{uv} \geq \alpha |I_t|.
\end{equation}

\begin{figure*}
     \centering
     \begin{subfigure}[b]{0.32\textwidth}
         \centering
         \includegraphics[width=\textwidth]{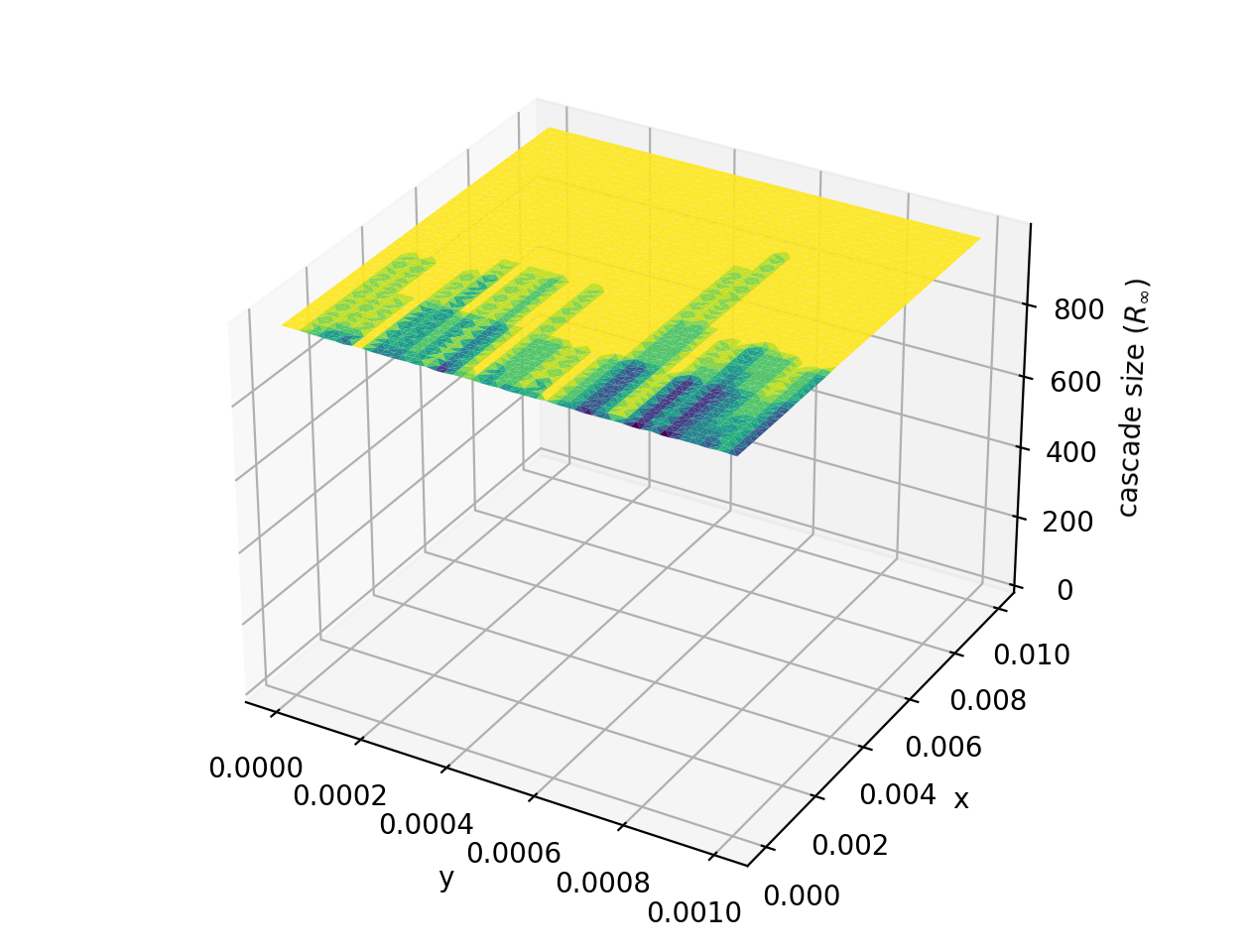}
         \caption{False Content, Balanced, 2 partitions, Control}
         \label{fig:w16}
     \end{subfigure}
     \hfill
     \begin{subfigure}[b]{0.32\textwidth}
         \centering
         \includegraphics[width=\textwidth]{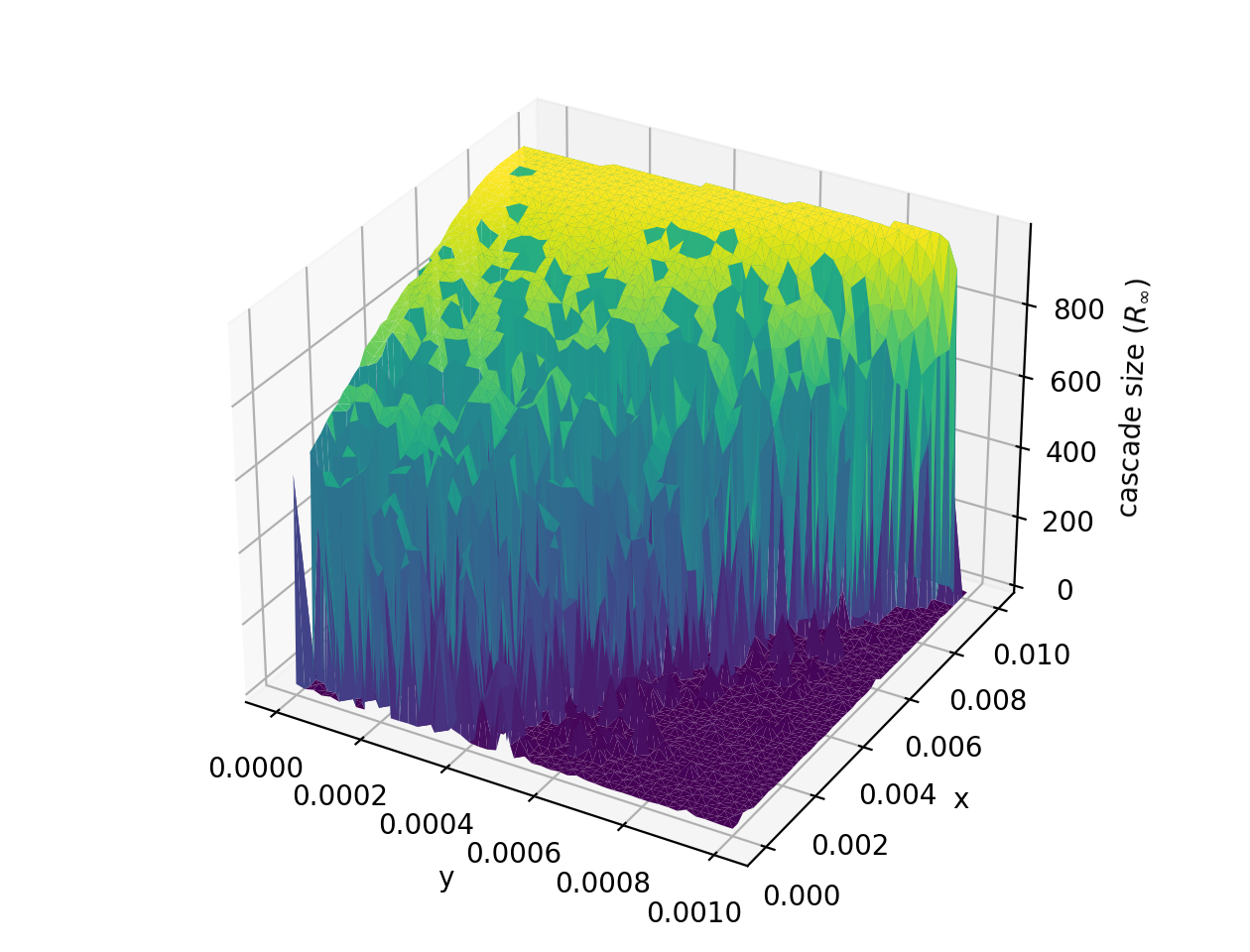}
         \caption{False Content, Balanced, 2 partitions, $\alpha = 1.5, \lambda = 1$}
         \label{fig:w26}
     \end{subfigure}
     \hfill
     \begin{subfigure}[b]{0.32\textwidth}
         \centering
         \includegraphics[width=\textwidth]{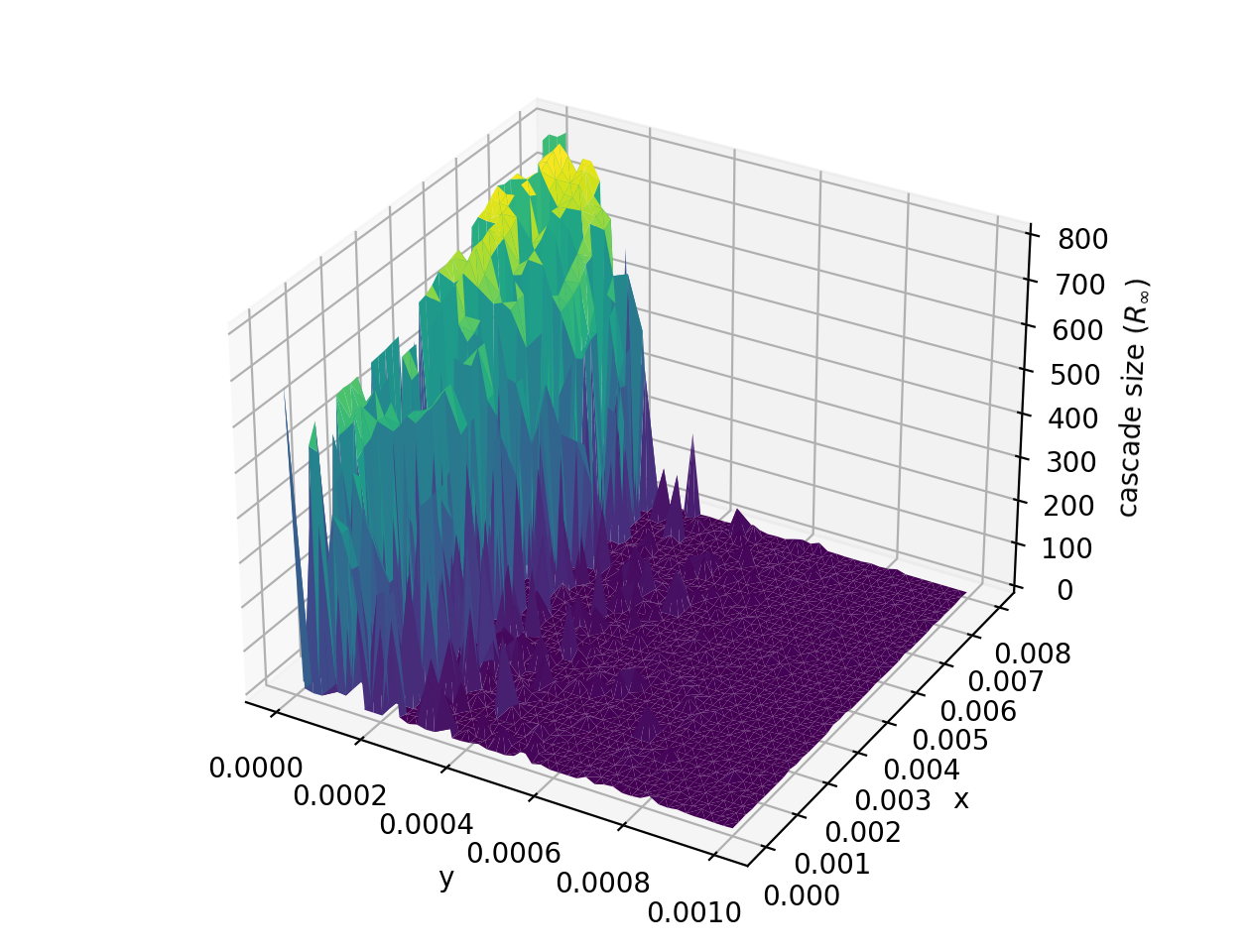}
         \caption{False Content, Unbalanced, 3 partitions, $\alpha = 1.5, \lambda = 1$}
         \label{fig:w36}
     \end{subfigure}
     \begin{subfigure}[b]{0.32\textwidth}
         \centering
         \includegraphics[width=\textwidth]{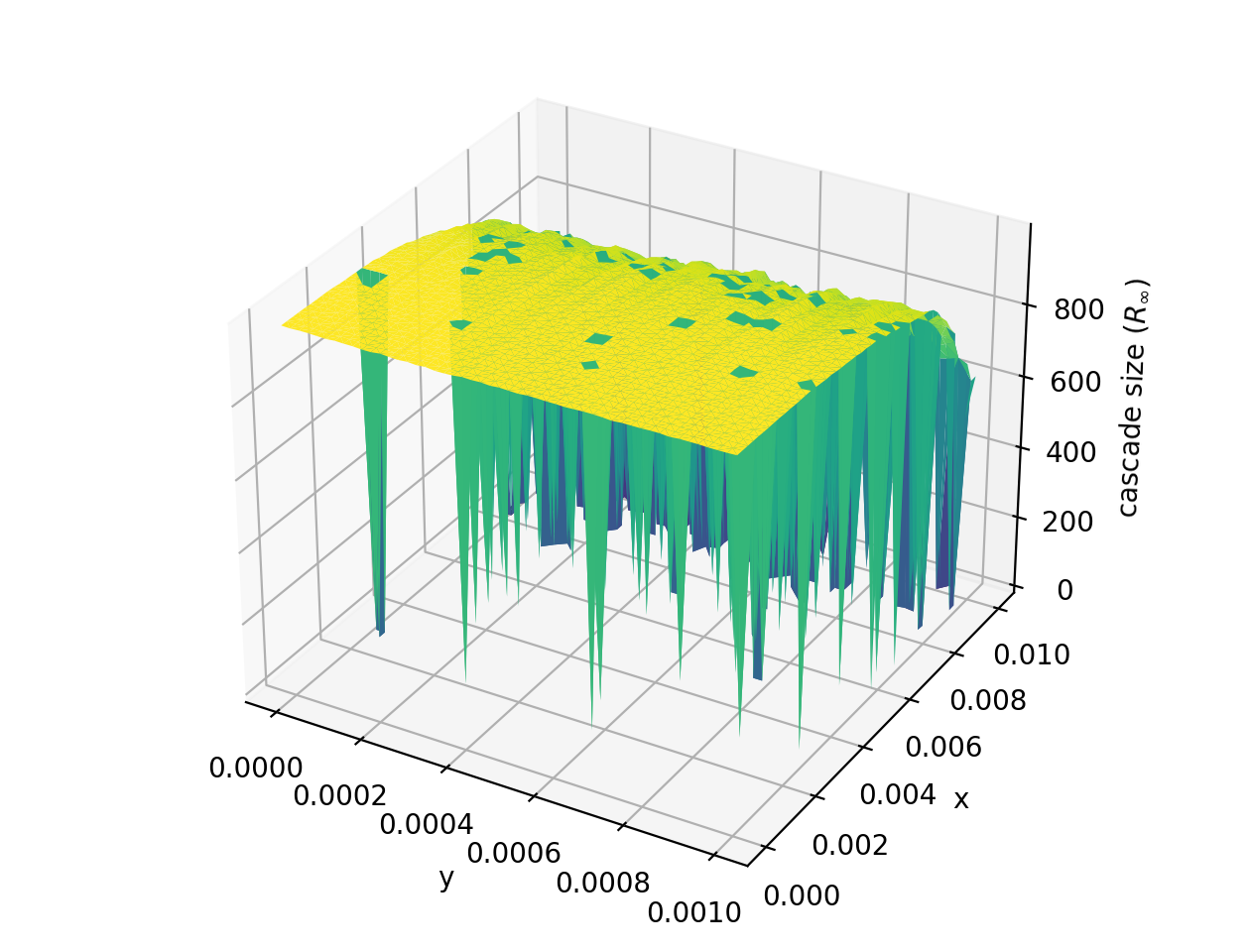}
         \caption{True Content, Balanced, 2 partitions, Control}
         \label{fig:w46}
     \end{subfigure}
     \hfill
     \begin{subfigure}[b]{0.32\textwidth}
         \centering
         \includegraphics[width=\textwidth]{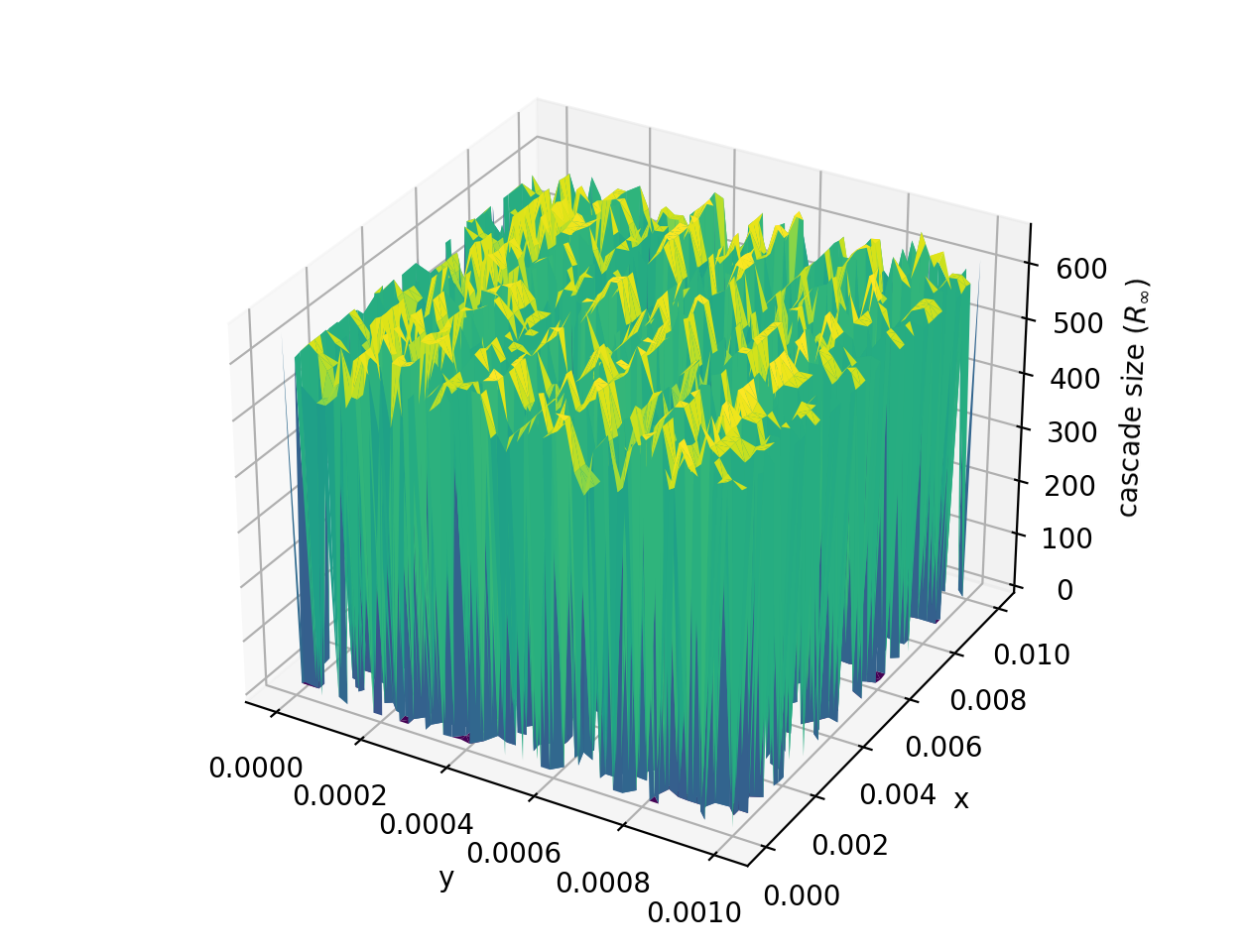}
         \caption{True Content, Balanced, 2 partitions $\alpha = 1.5, \lambda = 1$}
         \label{fig:w56}
     \end{subfigure}
     \hfill
     \begin{subfigure}[b]{0.32\textwidth}
         \centering
         \includegraphics[width=\textwidth]{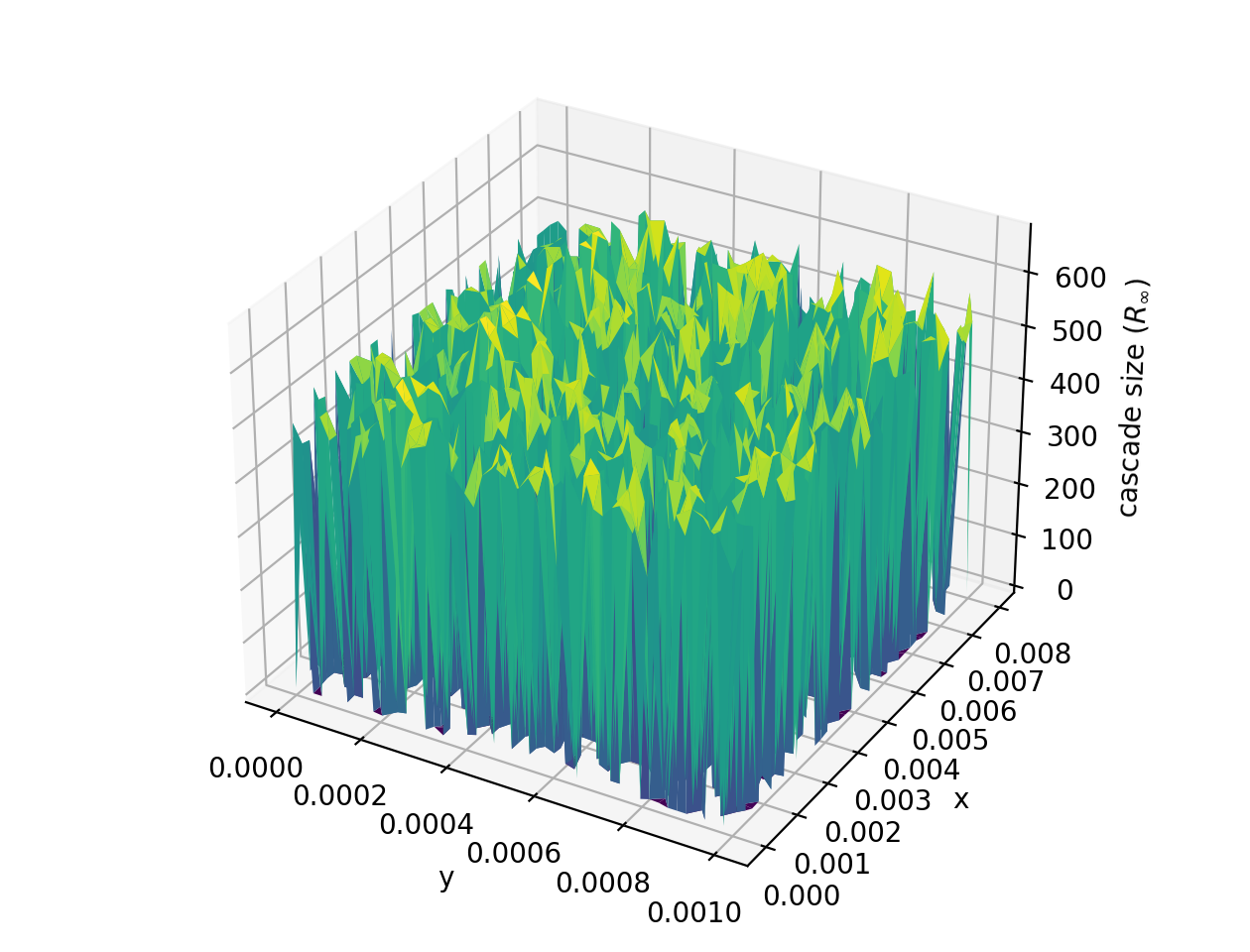}
         \caption{True Content, Unbalanced, 3 partitions $\alpha = 1.5, \lambda = 1$}
         \label{fig:w66}
     \end{subfigure}
        \caption {Cascade size of false and true content across two different SBM configurations. }
        \label{fig::WvsRegret}
\end{figure*}


In application, viral true content almost always satisfies this feasibility condition for some $\alpha \geq 1$ in the initial propagation period. However, less viral content may violate the feasibility condition. In this case, our original optimization goals cannot be reached. To counter this issue, for content that violates the feasibility criterion \eqref{feas_criterion} infeasible we can soften the linear program \eqref{eqn::dropout_problem_asymptotic_linear}. That is,
\begin{equation}
\label{eqn::dropout_problem_asymptotic_linear_soft}
    \min_{d \in [0,1]^{k\times k}} \quad \sum_{v = 1}^k \sum_{u = 1}^k |S^v_{t}| |I^u_t| d_{uv}b^-_{uv} + \lambda |S^v_{t}|  |I^u_t| d_{uv}b^+_{uv},
\end{equation}
where $\lambda$ is a weight parameter that signifies the importance of preserving true content relative to the importance of suppressing false content. 

We can use the linear programs \eqref{eqn::dropout_problem_asymptotic_linear} and \eqref{eqn::dropout_problem_asymptotic_linear_soft} in conjunction to provide a general approximate solution the our main problem \eqref{eqn::dropout_problem} given in section 3.4. The Algorithm \ref{alg::dropout} provides a method to achieve which this conjunction. Here $\opt(*)$ refers to the optimal solution on the optimization problem $*$, which in the case of Algorithm \ref{alg::dropout} can be found efficiently using existing linear program solution methods.

\begin{algorithm}
    \caption{False Content Minimization Using Dropouts}
    \begin{algorithmic}[1]
    \REQUIRE {Networks $\RandGraph^-, \RandGraph^+$, 
    Real-World Network $\RandGraph$, 
    Set of seed users $I_0$, 
    Safety parameter $\alpha$,
    Weight $\lambda$.}
    \STATE $t \gets 0$
    \STATE $S_0 \gets V \setminus I_0$
    \STATE $R_0 \gets \emptyset$
    \WHILE{$|I_t| > 0$}
    \IF{$\sum_{v = 1}^k \sum_{u = 1}^k |S^v_{t}|  |I^u_t|b^+_{uv} \geq \alpha |I_t|.$}
        \STATE $d^* = \opt( \textrm{linear program \eqref{eqn::dropout_problem_asymptotic_linear}})$
    \ELSE
        \STATE $d^* = \opt( \textrm{linear program \eqref{eqn::dropout_problem_asymptotic_linear_soft}})$
    \ENDIF
    \STATE $\altRandGraph \gets \alter_{\alterfun_{d^*}} (\RandGraph)$ \COMMENT {Alter the real-world social network.}
    \STATE $S_{t+1}, I_{t+1}, R_{t+1} \gets$ Observe $(S_{t+1}, I_{t+1}, R_{t+1}| S_t, I_t, R_t, \altRandGraph)$
    \STATE $t \gets t + 1$
    \ENDWHILE
    \end{algorithmic}
    \label{alg::dropout}
\end{algorithm}

\section{Experimental Results}

We test Algorithm \ref{alg::dropout} both on synthetic stochastic block model networks and on a real misinformation dataset collected over Twitter. The dataset we used is called WICO \cite{wico} which contains over $3500$ separate tweets and status updates collected between January 2020 and July 2020. In these tests, we use the total cascade size $R_\infty$, which is the total size of the set of users that have received a piece of content after the SIR propagation terminates, as the performance metric. For the case of true content, we want $R_\infty$ to be high, and for the case of false content, we want $R_\infty$ to be low.

\subsection{Experiments Using Synthetic Model }
In this section, we test the performance of Algorithm \ref{alg::dropout} using synthetic social networks that are modeled as SBMs. We test the effectiveness of Algorithm \ref{alg::dropout} over four different test configurations. We name these configurations as follows: Balanced with $2$ partitions, unbalanced with $2$ partitions, balanced with $3$ partitions, and unbalanced with $3$ partitions. All of these configurations have $1000$ users. The balanced configurations have partition sizes of $[500,500]$ for $2$ partition case and $[334,333,333]$ for $3$ partition case. The unbalanced configurations have partition sizes of $[800,200]$ for $2$ partition case and $[500,300,200]$ for $3$ partition case.

We associate a base matrix $\mathbf{b}_{\textrm{base}}$ with each of these test configurations. For configurations with $2$ partitions this base matrix is defined as,
\begin{equation}
    \mathbf{b}_{\textrm{base}} \defeq \begin{bmatrix}
        0.01 & 0.002 \\
        0.002 & 0.01
    \end{bmatrix}.
\end{equation}
Similarly, for configurations with $3$ partitions, we define this base matrix as,
\begin{equation}
    \mathbf{b}_{\textrm{base}} \defeq \begin{bmatrix}
        0.01 & 0.002& 0.002 \\
        0.002 & 0.01& 0.002 \\
        0.002 & 0.002& 0.01 \\
    \end{bmatrix}.
\end{equation}

We use these base matrices to generate SBMs that simulate true and false content propagation. To test the effect of different content propagation dynamics on the performance of algorithm \ref{alg::dropout}, we select two parameters $x \in [0, 0.01]$ and $y \in [0, 0.001]$. For each configuration, we sweep across the range of possible $x$ and $y$ combinations with $50$ subdivisions in each dimension. For each $x$ and $y$ choice, we generate the SBMs $\RandGraph^+$ and $\RandGraph^-$ that describe the content transfer probabilities for true and false content respectively. We define the SBM matrices for $\RandGraph^+$ and $\RandGraph^-$ as follows: For true content we define
\begin{equation}
    [b^+_{uv}]_{uv} = \mathbf{b}_{\textrm{base}} + x\mathbb{I} - y (\mathbb{J} - \mathbb{I}),
\end{equation}
and for false content we define
\begin{equation}
    [b^-_{uv}]_{uv} = \mathbf{b}_{\textrm{base}} - x\mathbb{I} + y (\mathbb{J} - \mathbb{I}),
\end{equation}
where $\mathbb{I}$ is the identity matrix and $\mathbb{J}$ is the all-ones matrix. Then we simulate content propagation over these networks and determine the cascade sizes $R_{\infty}$ that result from different choices for $\alpha$ and $\lambda$.

Table \ref{tab:syth}, summarizes the normalized mean cascade size $\expec[R_\infty]/N$ and ratio of tests that have cascade size less than $N / 10$ to all tests. These statistics are collated over the complete range of all $x$ and $y$ combinations. For all configurations, we test two different parameter assignments for $\alpha$ and $\lambda$. The rows indicated as $(\alpha, \lambda) = (-,-)$ are control groups with no network alterations. This table shows that regardless of the choice of parameters $\alpha$ and $\lambda$, on average Algorithm \ref{alg::dropout} manages to reduce false content more than it reduces true content. Moreover, $\prob[R_\infty < {N / 10}]$ are much higher on false content compared to true content. The fact that $\prob[R_\infty < {N / 10}]$ are high on false content indicates that the cascade size $R_\infty$ has high variance. That is, the performance of Algorithm \ref{alg::dropout} varies greatly depending on the dynamics of the social network structure.

\begin{table*}[ht]
\centering
  \caption{Summmary of Results for Synthetic Tests}
  \label{tab:syth}
  \begin{tabular}{@{}cc|cc|cc|cc}
    \toprule
    SBM Type& Partitions & $\alpha$& $\lambda$&\multicolumn{2}{c}{Mean Cascade Size ($\expec[R_\infty]/N$)} & \multicolumn{2}{c}{Low Cascades ($\prob[R_\infty < {N / 10}]$)}\\
    & & && True Content & False Content & True Content & False Content \\
    \midrule
    Balanced& 2& -& -& 0.89 & 0.98& 0.08&  0.09\\
    Balanced& 2& 1.5& 1 & 0.51& 0.32& 0.28& 0.43\\
    Balanced& 2& 2& 1.5 & 0.72& 0.41& 0.18&  0.30\\
    Unbalanced& 2& -& - & 0.92& 0.96& 0.08&  0.08\\
    Unbalanced& 2& 1.5& 1 & 0.46& \textbf{0.14}& 0.22& \textbf{0.82}\\
    Unbalanced& 2& 2& 1.5& 0.68& 0.38& 0.27&  0.57\\
    Balanced& 3& -& -& 0.86 & 0.95& 0.12&  0.09\\
    Balanced& 3& 1.5& 1& 0.52 & 0.36& 0.33&0.48\\
    Balanced& 3& 2& 1.5& 0.73 & 0.48& 0.18&0.37\\
    Unbalanced& 3& -& -& 0.88 & 0.97 & 0.11&  0.13\\
    Unbalanced& 3& 1.5& 1& 0.51 & 0.23 &0.29& 0.70\\
    Unbalanced& 3& 2& 1.5& 0.66 & 0.37 &0.22& 0.64\\
  \bottomrule
\end{tabular}
\end{table*}

Figure \ref{fig::WvsRegret} shows the average cascade size three configurations over the full span of $x, y$ combinations, where the average is computed over $50$ separate trials. This figure shows that the Algorithm \ref{alg::dropout} affects the true content in a similar way across all $x, y$ values. This is of course expected since the constraint in the linear program  \eqref{eqn::dropout_problem_asymptotic_linear} in Algorithm \ref{alg::dropout} fixes the expected propagation rate of true content. On the contrary, the effect of Algorithm \ref{alg::dropout} on the false content depends heavily on both the value of $x$ and $y$ and the overall structure of the social media network. In general, there is a sharp boundary transition in the cascade size of the false content and the Algorithm \ref{alg::dropout} tends to either reduce the cascade size of false content to near $0$, or have very little impact to the propagation of the false content. This explains the high  $\prob[R_\infty < {N / 10}]$ values seen in Table \ref{tab:syth}. The sharp transition in cascade size seen in Figures \ref{fig:w26} and \ref{fig:w36} is caused by the same mechanism that causes the state-transition-like behavior that is present in most complex real networks, where a giant connected component can appear suddenly as we increase the overall expected degree of nodes in a large graph \cite{bollobas1998random}.

\subsection{ Experiments Using Real World Data}
We use a pre-existing dataset named WICO \cite{wico} for these tests. This dataset contains share times and propagation networks for separate pieces of content. These content are labeled as follows:
\begin{enumerate}
    \item 5G-Corona Conspiracy: Conspiracy content that claims there is a causation between the Covid-19 pandemic and 5G,
    \item Other Conspiracy,
    \item Non-Conspiracy.
\end{enumerate}
In this dataset, we re-label content that is labeled as ``5G-Corona Conspiracy'' or ``Other Conspiracy'' as false content, and we re-label Non-Conspiracy content as true content. 

We then assign polarizations to each user by running  modularity based clustering \citep{blondel2008fast} with resolution \citep{lambiotte2008laplacian} set to $2$ on the union of all graphs in the WICO dataset. The resulting network has $153779$ nodes and $216848$ edges. The modularity class assignment has $63914$ partitions. We restrict the number of partitions by merging all partitions with a number of users less than $1\%$ of the total number of users in the merged graph. The resulting partitioning of the graph has $13$ partitions. These partitions correspond to different echo chambers in the network, therefore they are a close approximation for the polarization classes of the users in the social network.

After determining polarization classes, we fit the model networks $\RandGraph^+$ and $\RandGraph^-$ to the dataset by frequentist estimation of SBM matrices $[b^+_{uv}]_{uv}$ and $[b^-_{uv}]_{uv}$ by counting the number of content transfers between different polarization groups. We then use Algorithm \ref{alg::dropout} to generate dropout-based alterations on the actual social media network. We simulate the content propagation under these dropout-based alterations by sampling a random content from the dataset and then following its propagation while randomly dropping content transfers based on dropout probabilities $d^*$ given by Algorithm \ref{alg::dropout}. 

We test the three different parameter settings for Algorithm \ref{alg::dropout}. These settings are $(\alpha, \lambda) = (1.5, 1)$, $(\alpha, \lambda) = (2, 1.5)$, $(\alpha, \lambda) = (3, 2)$. Table \ref{tab:real} shows the resulting cascade size statistics, averaged over $500$ samples, for these each of these settings as well as a control group which is denoted as $(\alpha, \lambda) = (-, -)$. Contrary to the previous synthetic tests, we do not normalize the expected cascade size value in Table \ref{tab:real}, since the cascade sizes of these networks are very small compared to the total number of users in the network. The fact that the cascade sizes are small is not surprising, since often in the real world social media network only a small fraction of users tend to participate in re-sharing a piece of content they receive due to the vastness of the number of available content and the variability of interests of users. The average performance of Algorithm \ref{alg::dropout} decreases in these real-world datasets compared to synthetic model test due to the inaccuracies in the SBM models $\RandGraph^+$ and $\RandGraph^-$. However, for all choices of $(\alpha, \lambda)$ Algorithm \ref{alg::dropout} achieves discrimination between true and false content.

\begin{table}[h]
\centering
  \caption{Summmary of Results for WICO Dataset Tests}
  \label{tab:real}
  \begin{tabular}{cc|cc|cc}
    \toprule
    $\alpha$& $\lambda$&\multicolumn{2}{c}{$\expec[R_\infty]$} & \multicolumn{2}{c}{$\prob[R_\infty < 5$]}\\
     & & True C. & False C.& True C.& False C.\\
    \midrule
    -& -& 50.1 & 48.7& 0.01&  0.00\\
    1.5& 1 & 32.8& 26.1& 0.05& 0.13\\
    2& 1.5 & 39.4& 28.2& 0.04&  0.09\\
    3& 2 & 41.1& 36.0& 0.00& 0.02 \\
  \bottomrule
\end{tabular}
\end{table}

\section{Conclusion and Future Work}
We demonstrated that it is possible to counter misinformation without explicit identification of misinformation content by altering the content propagation dynamics on social network. A major advantage of our approach is that it does not require the system to be able to identify if a particular news item contains misinformation content or not. Furthermore, our approach can be used in conjunction with these detection algorithms to improve the effectiveness of misinformation control while maintaining some of the advantages offered by the network-design-based approach. In our future studies, we will investigate this possibility further.

Throughout this work we have assumed that content is either true and false. In reality this clear of a separation between true and false content types is rarely possible. It is possible to extend our methods and algorithms to admit more content types, which can increase the performance of Algorithm \ref{alg::general} since more content types can achieve a more nuanced description of real misinformation dynamics. However, this extension requires modifications on the problem statement and the formulation of the problem. 

\section*{Acknowledgements}
This work was supported in part by ONR N00014-21-1-2502 and NSF 1652113.

\Urlmuskip=0mu plus 1mu\relax
\bibliography{ref}

\end{document}